\documentclass[ notitlepage,amsmath,amssymb,preprintnumbers,superscriptaddress,nofootinbib,aps,twocolumn,prl,floatfix]{revtex4-2}

\usepackage{hyperref}
\usepackage{xcolor}
\usepackage{graphicx}
\usepackage{braket}
\usepackage{comment}
\usepackage{slashed}

\newcommand{\nn}{\nonumber\\}
\newcommand{\tr}{\mathrm{Tr}}
\newcommand{\ie}{\textit{i.e.}}
\newcommand{\factorAlpha}{\alpha^{z_1,z_2}_{M_1,M_2}}

\usepackage[normalem]{ulem}

\begin{abstract}
Spin correlations between particles produced at colliders provide valuable insights for quantum information studies. While traditional studies of quantum information at colliders are typically limited to massive particles with perturbative decay, we propose an innovative method to explore the Bell inequality in massless quark pair systems by analyzing the azimuthal correlations in $\pi^+\pi^-$ dihadron pair production at lepton colliders. Revisiting the Belle data, we have shown the potential to detect Bell inequality violation of light quarks by introducing an additional angular cut, achieving a significance of 2.5 $\sigma$ even in the worst-case scenario of 100\% correlated systematic uncertainties in each bins. The significance substantially exceeds $5\sigma$ when considering uncorrelated systematic uncertainties. Our approach opens avenues for exploring spin quantum information with non-perturbative processes as spin analyzer and leverages existing data for quantum information research.
\end{abstract}

\begin{document}
\preprint{PITT-PACC-2414, CPTNP-2025-020}

\title{Bell Inequality Violation of Light Quarks in Dihadron Pair Production at Lepton Colliders}
\author{Kun Cheng}
\email{kun.cheng@pitt.edu}
\affiliation{PITT PACC, Department of Physics and Astronomy,\\ University of Pittsburgh, 3941 O’Hara St., Pittsburgh, PA 15260, USA}

\author{Bin Yan}
\email{yanbin@ihep.ac.cn (corresponding author)}
\affiliation{Institute of High Energy Physics, Chinese Academy of Sciences, Beijing 100049, China}
\affiliation{Center for High Energy Physics, Peking University, Beijing 100871, China}

\date{\today}

\maketitle

\emph{Introduction.---}
Quantum entanglement and Bell inequality violations are distinctive features of quantum systems that distinguish them from classical systems. In recent years, colliders have emerged as a promising environment for studying these quantum phenomena.  Notably, there has been extensive exploration of quantum information in top quark pair production at colliders ~\cite{Afik:2020onf,Fabbrichesi:2021npl,Severi:2021cnj,Afik:2022kwm,Aguilar-Saavedra:2022uye,Afik:2022dgh,Dong:2023xiw,Han:2023fci,Maltoni:2024csn,White:2024nuc,Han:2024ugl}.
The significantly shorter lifetime of the top quark compared to the timescale of hadronization allows the top quark to decay electroweakly before hadronization. Consequently, the spin information of a top quark is transferred to its decay products, enabling a perturbative reconstruction of the top quark's spin density matrix from its decay distributions. This traditional method of quantum tomography at colliders, known as the ``decay approach"~\cite{Cheng:2024rxi}, has resulted in the detection of quantum entanglement in top pair events as reported by both the ATLAS~\cite{ATLAS:2023fsd} and CMS~\cite{CMS:2024pts} collaborations. 
This methodology has also been extensively utilized for other massive particle pairs, such as $\tau$ lepton pairs~\cite{Altakach:2022ywa,Ehataht:2023zzt,Fabbrichesi:2024wcd,Du:2024sly}, massive gauge boson pairs~\cite{Barr:2022wyq,Ashby-Pickering:2022umy,Aguilar-Saavedra:2022wam,Fabbrichesi:2023cev,Fabbri:2023ncz,Bi:2023uop}, $Y\bar Y$ pairs~\cite{Wu:2024asu} and $\Lambda_b\bar\Lambda_b$ pairs~\cite{Du:2024sly,Afik:2024uif}.

While the decay approach is only valid for massive particle pairs that decay perturbatively, exploring spin entanglement and Bell inequality violations in light quark pairs ($q\bar q$) presents a compelling research avenue that has been overlooked in the literature. In contrast to the top quark, the light quark does not decay but instead fragments into a jet of hadrons after produced from hard scattering. Nevertheless, the spin of a light quark can still be transferred to its hadron final states and can be systematically characterized by the non-perturbative fragmentation functions. Within the framework of collinear factorization, it has been demonstrated that only the transverse spin of a light quark can be transferred to the unpolarized hadron final states, and reconstructing it requires the presence of at least two hadrons $h_a(\vec{p}_a)$ and $h_b(\vec{p}_b)$ within the jet~\cite{Bianconi:1999cd}. The transverse spin $\vec{s}_T$ of a quark disrupts the rotational symmetry around its momentum direction, leading to an azimuthal asymmetry in the jet constituents. This asymmetry is not accounted for in perturbative calculations but captured by interference dihadron fragmentation functions (diFFs)~\cite{Bianconi:1999cd,Barone:2001sp,Zhou:2011ba,Metz:2016swz,Wen:2024cfu,Huang:2024awn,Wen:2024nff,Yang:2024kjn}~\footnote{The linear polarization of gluons could result in a similar azimuthal asymmetry~\cite{Yu:2022kcj,Guo:2024jch}.}. Thus, the dihadron pair production $e^+e^-\to q\bar{q}\to (h_{a1},h_{b1})+(h_{a2},h_{b2})+X$ offers a promising avenue for investigating the transverse spin correlation of $q\bar q$ produced at lepton colliders~\cite{Artru:1995zu,Boer:2003ya}.

The massless quark pair system also manifests a significant advantage for Bell inequality studies, as the transverse spins of $q$ and $\bar q$ generated from the $e^+e^-\to \gamma^*\to q\bar q $ processes are 100\% correlated in the central scattering region. The corresponding spin state forms a maximally entangled Bell state $\frac{1}{\sqrt{2}}(\ket{\uparrow_{y}\downarrow_{y}}+\ket{\downarrow_{y}\uparrow_{y}})$, with $\ket{\uparrow_{y}}$ and $\ket{\downarrow_{y}}$ denoting the transverse polarized eigenstates along $\hat{y}$, the normal direction of the hard scattering plane. Therefore, although the complete $q\bar q$ spin density matrix, encompassing longitudinal spin, cannot be reconstructed from dihadron pairs in the collinear limit, the transverse spin already presents the most relevant information about the $q\bar q$ spin Bell state.
By applying a simple selection cut on the hard scattering angle, an ideal ensemble of the spin Bell state can be prepared from the $e^+ e^-$ annihilation.  The maximal entangled behavior makes the $q\bar q$ system attractive for investigating quantum information, especially, offering the potential for a distinct signal of Bell inequality violation that has not yet been observed at colliders using the decay approach.

In this Letter, we propose a novel approach to perform a partial quantum tomography of $q\bar q$ spin state by utilizing dihadron interference fragmentation functions as spin analyzing powers to reconstruct the transverse parts in the $q\bar q$ spin density matrix, and present a sufficient condition for detecting entanglement and Bell inequality violation in $q\bar q$ production at lepton colliders. Our approach represents the first exploration of Bell inequality violation within the massless quark pair system, opening up avenues for spin quantum information research with non-perturbative processes as spin analyzer.

With the method we have developed, numerous existing data becomes readily available for quantum information studies. As a first example, we revisit the $\pi^+\pi^-$ dihadron pair production at the $10.58$\,GeV Belle collider~\cite{Belle:2011cur}. Our analysis indicates that the current uncertainty implies a potential observation of the Bell inequality violation in the $q\bar q$ system from existing data, with a significance of at least $2.5\,\sigma$. A more detailed examination of systematic uncertainties is expected to yield an even much larger significance, possibly resulting in the first measurement of Bell inequality violation in a collider scattering process.

\begin{figure}
    \centering
    \includegraphics[width=\linewidth]{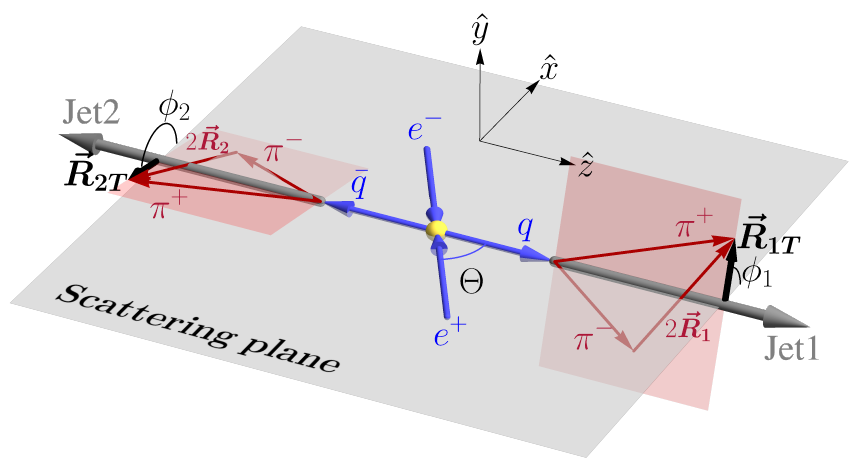}
    \caption{Leading order kinematic configuration of $\pi^+\pi^-$ dihadron pair production at lepton colliders.}
    \label{fig:geometry}
\end{figure}

\emph{Spin state of quark pair and the Bell inequality.---}
We start with the most general spin density matrix of $q\bar q$ system, which is parametrized as
\begin{equation}\label{eq:rhoInCij}
    \rho = \frac{I_2\otimes I_2 + B_i \sigma_i\otimes I_2 + \bar{B}_i I_2 \otimes \sigma_i + C_{ij} \sigma_i\otimes \sigma_j }{4},
\end{equation}
where $I_2$ represents the 2-dimensional identity matrix and $\sigma_i$ denotes the Pauli matrix.  Then the spin density matrix is fully characterized by the net polarization vectors $B_i$ ($\bar B_i$) of $q$ ($\bar q$), and the spin correlation $C_{ij}$ between $q$ and $\bar q$.  For the massless quarks, it is convenient to choose the helicity basis in the rest frame of $q\bar q$ system, where the $\hat{z}$ direction is chosen as the quark momentum direction and the spin components of $q\bar  q$ inside the $\hat{x}$-$\hat{y}$ plane represent their transverse spins; see Fig.~\ref{fig:geometry} for the coordinate.\footnote{The same coordinate frame is often referred to as $(\hat r,\hat n,\hat k)$ when investigating the quantum information of top quark pairs~\cite{Afik:2020onf,Fabbrichesi:2021npl}.}

For a local theory, the Clauser-Horne-Shimony-Holt (CHSH) type Bell inequality is expressed as follows~\cite{Clauser:1969ny}
\begin{equation}
|
\vec a_1\cdot \mathbf{C} \cdot \vec b_1+
\vec a_1\cdot \mathbf{C} \cdot \vec b_2+
\vec a_2\cdot \mathbf{C} \cdot \vec b_1-
\vec a_2\cdot \mathbf{C} \cdot \vec b_2
|\leq 2,
\end{equation}
where $\vec a_i$ and $\vec b_i$ represent normalized directions to measure the spin of $q$ and $\bar q$, respectively, and $\mathbf{C}$ is the correlation matrix $C_{ij}$ in Eq.~\eqref{eq:rhoInCij}. By scanning the four spin measurement directions, the violation of the CHSH inequality occurs if and only if the two largest eigenvalues $\mu_{1,2}$ of $\mathbf{C}^T\mathbf{C}$ satisfy $\mu_1+\mu_2>1$~\cite{HORODECKI1995340}.
In practice, it is convenient to choose $\vec a_i$ and $\vec b_i$ from the axis directions $\hat x,\hat y$ and their diagonal directions~\cite{Severi:2021cnj}, then utilize the linear combinations of the correlation matrix,
\begin{equation}\label{eq:BpmDef}
    \mathcal{B}_{\pm} \equiv C_{xx}\pm C_{yy},
\end{equation}
to test the CHSH inequality violation. A sufficient condition for the Bell inequality violation is
\begin{equation}
    |\mathcal{B}|>\sqrt{2}
\end{equation}
where the Bell variable $\mathcal{B}$ can be either $\mathcal{B}_+$ or $\mathcal{B}_-$.
The maximal value of the Bell variable in quantum theory is $\mathcal{B}=2$, which is achieved when the state is a maximally entangled Bell state.
In the following we only focus on the transverse parts $C_{xx}$ and $C_{yy}$ of the spin correlation of $q\bar q$ as they are sufficient to measure Bell inequality violation.

In the Standard Model (SM), the spin correlation matrix of $q\bar q$ in the $e^- e^+ \to q\bar q$ process  at Belle is
\begin{equation}\label{eq:Cij}
C_{ij} = \mathrm{diag}\left( \frac{\sin^2\Theta}{1+\cos^2\Theta},-\frac{\sin^2\Theta}{1+\cos^2\Theta},1 \right)
\end{equation}
where $\Theta$ represents the hard scattering angle,
and the net polarization vectors of $q$ and $\bar q$ are zero, \ie, $B_i = \bar{B}_i=0$.
We see that the transverse spin of $q$ and $\bar q$ along $x$($y$) direction are positively(negatively) correlated~\cite{Chen:1994ar}, both achieving 100\% correlation when $\Theta=\pi/2$. Then the Bell variables are
\begin{equation}\label{eq:SMprediction}
   \mathcal{B}_+=0,~~ \mathcal{B}_{-}=\frac{2\sin^2\Theta}{1+\cos^2\Theta}.
\end{equation}

In the central scattering region, ideally with $\cos\Theta=0$, the correlation matrix $C_{ij}=\mathrm{diag}(1,-1,1)$ corresponds to the spin triplet Bell state $\frac{1}{\sqrt{2}}(\ket{\uparrow_{y}\downarrow_{y}}+\ket{\downarrow_{y}\uparrow_{y}})$, and $\mathcal{B}_-$ achieves its maximum value $\mathcal{B}^{\rm max}=2$. 
In practice, a finite angular selection
\begin{equation}
|\cos\Theta|<c_{\rm max}
\end{equation}
is required to collect events. Gathering the $q\bar q$ event produced at different scattering angles results in an angular-averaged state, defined as
\begin{equation}
    \bar\rho = \frac{1}{\sigma_{\rm hard}}\int_{-c_{\rm max}}^{c_{\rm max}} \frac{\mathrm{d}\sigma_{\rm hard}}{\mathrm{d}\cos\Theta} \rho(\Theta) ~ \mathrm{d}\cos\Theta,
\end{equation}
where $\rho(\Theta)$ is the density matrix of the $q\bar q$ produced at a fixed scattering angle $\Theta$ with azimuthal angle of $q\bar q$ averaged over, and $\sigma_{\rm hard}$ is the hard scattering cross section within the angular cut.
The angular-averaged state is also known as ``fictitious state"~\cite{Afik:2022kwm} due to introducing the subtleties of basis dependence.  
However, for massless quarks, the conventional choice of helicity basis is already optimal for testing Bell inequality violation~\cite{Cheng:2023qmz,Cheng:2024btk}. 

Including $q\bar q$ that are not produced in the central angle region results in a mixed state rather than a pure Bell state, which dilutes the signal of Bell inequality violation. This mixture can be characterized by the purity $\mathrm{Tr}(\bar\rho^2)$ of the state $\bar \rho$, where $\mathrm{Tr}(\bar\rho^2)=1$ if $\bar \rho$ is a pure state and $\mathrm{Tr}(\bar\rho^2)<1$ for mixed states. To select the maximally entangled region, a relatively narrow selection cut can be chosen; \textit{e.g.}, the events collected with $c_{\rm max}=0.1$ approximately correspond to the Bell state with $\mathrm{Tr}(\bar\rho^2)\simeq 0.99$.

\emph{Fragmentation of quark pair spin state.---}
We consider the semi-inclusive process $e^+ e^- \to q (k_1)\bar q (k_2)\to (\pi^+\pi^-)+(\pi^+\pi^-)+X$, with the leading-order kinematics depicted in Fig.~\ref{fig:geometry}.  The kinematics of a $\pi^+\pi^-$ pair are characterized by their total momentum $P_{i}=p_i^{\pi^+}+p_i^{\pi^-}$ (for $i=1,2$) and their momentum difference $R_i\equiv (p_i^{\pi^+}-p_i^{\pi^-})/2$.  We consider the collinear limit where the total momentum $\vec P_i$ is aligned with the quark momentum $\vec k_i$ and approximately the jet directions, and the dihadron fragmentation phase space in this limit is conventionally described by the dihadron invariant mass $M_i\equiv P_i^2$, the azimuthal angle $\phi_i$ of the relative transverse momentum $\vec{R}_{i,T}$, and the total momentum fractions $z_1=P_1^+/k_1^+$ and $z_2=P_2^-/k_2^-$ where $P_i^\pm (k_i^\pm)$ represent the light-cone components of the momentum of $P_i (k_i)$.

Under the collinear factorization ($|\mathbf{P}_i|\gg M_i$), the differential distribution of $\pi^+ \pi^-$ pairs fragmented from a quark pair with spin density matrix $\bar\rho_{ss',\bar s \bar s'}$ can be written as~\cite{Collins:1993kq,Barone:2001sp,Collins:2011zzd}:
\begin{equation}\label{eq:density}
    \frac{\mathrm{d}\sigma}{\mathrm{d}\Omega_1 \mathrm{d}\Omega_2} = \sigma_{\rm hard} \sum_{ss'\bar{s}\bar{s}'} \bar\rho_{ss',\bar s\bar s'}\mathcal{D}^{ss'}_{\pi^+\pi^-/q} \mathcal{D}^{\bar s\bar s'}_{\pi^+\pi^-/\bar q},
\end{equation}
where $\mathrm{d}\Omega_{i}=\mathrm{d}z_i \mathrm{d}M_i \mathrm{d}\phi_i $ represents the phase space of the $\pi^+ \pi^-$ pair fragmented from $q$ or $\bar q$,  $(s,s',\bar s, \bar{s}')$ are spin indices of $q$ and $\bar q$,  $\mathcal{D}_{\pi^+\pi^-/q}^{ss'}$ represents the dihadron fragmentation matrix of the parton $q$ into a $\pi^+\pi^-$ pair, and $\sigma_{\rm hard}$ is the hard scattering cross section within the angular selection.  The scattering process is therefore factorized into two components: the hard parts $\sigma_{\rm hard}$ and $\bar\rho_{ss',\bar{s}\bar{s}'}$ that can be calculated perturbatively, \textit{e.g.}, the SM predictions in Eq.~\eqref{eq:Cij}, and the collinear fragmentation part $\mathcal{D}_{\pi^+\pi^-/q}$ that can only be determined from global fittings with experimental data.  For unpolarized final-state hadrons, the parity reflection symmetry results in only two independent collinear factors for each flavor of quarks:
\begin{align}
&\frac{1}{2}\tr(\mathcal{D}_{\pi^+\pi^-/q})=  D_1^q\left(z_1, M_1\right), \\
&\frac{1}{2}\tr(\sigma_z \mathcal{D}_{\pi^+\pi^-/q}) = 0, \\
&\frac{1}{2}\tr(\sigma_i \mathcal{D}_{\pi^+\pi^-/q}) =-\frac{\varepsilon_T^{i j} R_{1,T}^j}{|\vec R_{1,T}|} H_1^{\sphericalangle,q}\left(z_1,M_1\right),
\end{align}
where $R_{1,T}^i $ denotes the transverse component of $R_1$ with $i=x,y$. The anti-quark fragmentation matrix $\mathcal{D}_{\pi^+\pi^-/\bar{q}}$ is related to $\mathcal{D}_{\pi^+\pi^-/q}$ through charge conjugation symmetry. Here, $D_1(z_1,M_1)$ is referred to as the unpolarized diFF, while $H_1^{\sphericalangle}(z_1,M_1)$ represents the interference diFF. In the framework of collinear factorization, no Lorentz structure is allowed for the $z$ component of the fragmentation matrix, which prevents the reconstruction of the longitudinal spin of quarks. However, longitudinal spin information is retained when considering general transverse momentum-dependent (TMD) fragmentation functions~\cite{Metz:2016swz}.

For a $q\bar q$ spin state characterized by Eq.~\eqref{eq:rhoInCij} with $B_i=\bar{B}_i=0$, we can express the cross section of Eq.~\eqref{eq:density} in terms of the correlation matrix $C_{ij}$,
\begin{widetext}
\begin{align}\label{eq:fullDistribution}
    \frac{\mathrm{d} \sigma}{ \mathrm{d}z_1\mathrm{d}z_2 \mathrm{d}M_1 \mathrm{d}M_2  \mathrm{d}\phi_1\mathrm{d}\phi_2}
    = \sigma_{\rm hard} &\Bigg[  \sum_q e_q^2 D_1^{q}(z_1,M_1)D_1^{\bar q}(z_2,M_2) \nn
    &+   \frac{1}{2} \sum_q e_q^2  H_1^{\sphericalangle,q}(z_1, M_1) H_1^{\sphericalangle,\bar q}(z_2, M_2) \Big(\mathcal{B}_- \cos(\phi_1+\phi_2)-\mathcal{B}_+ \cos(\phi_1-\phi_2)\Big)\Bigg], 
\end{align}
\end{widetext}
where $\mathcal{B}_{\pm}=C_{xx}\pm C_{yy}$ are the Bell variables defined in Eq.\eqref{eq:BpmDef}, and $e_q$ denotes the electric charge of the quark $q$. Note that $CP$ symmetry is assumed so that there is no sine terms in the distribution.  With the SM predictions of $\mathcal{B}_\pm$ in Eq.~\eqref{eq:SMprediction}, Eq.~\eqref{eq:fullDistribution} reproduces the well-known Artru-Collins asymmetry~\cite{Artru:1995zu,Boer:2003ya}.

It has been shown that only the spin information of $u$- and $d$-quarks can be transferred to $\pi^+\pi^-$ pairs, while $H_1^{\sphericalangle,s}=H_1^{\sphericalangle,c}=0$ due to charge conjugation symmetry~\cite{Cocuzza:2023vqs}.  This implies that the $s$- and $c$-quarks solely contribute to the total cross section and dilute the asymmetry.  However, to facilitate comparison with the current experimental data, we will sum over all four flavors $q=u,d,s,c$ in the following analysis.  

Compared to the azimuthal correlation in the decay of top pairs~\cite{Aguilar-Saavedra:2022uye,Han:2023fci}, we observe that the ratio between $H_1^{\sphericalangle,q}$ and $D_1^q$ plays the role of an \textit{event-by-event} spin analyzing power.
The Bell variable $\mathcal{B}_+$ and $\mathcal{B}_-$ are, respectively, measured from the $\cos(\phi_1-\phi_2)$ modulation and $\cos(\phi_1+\phi_2)$ modulation, which is the same as the reconstruction of the Bell variable in the decay of top pairs~\cite{Aguilar-Saavedra:2022uye,Han:2023fci}.
Since the SM prediction for the Bell variable yields $\mathcal{B}_+=0$, we only focus on the variable $\mathcal{B}_-$, which is reconstructed by
\begin{align}\label{eq:BellVariable}
    \mathcal{B}_-&=  \frac{2\langle\cos(\phi_1+\phi_2)\rangle}{\alpha^{z_1,z_2}_{M_1,M_2}}  =  \frac{A_{12}}{\alpha^{z_1,z_2}_{M_1,M_2}}.
\end{align}
Here, $A_{12} \equiv  2 \langle \cos(\phi_1+\phi_2) \rangle$ represents the experimentally observed azimuthal asymmetry, measured from the average value of $\cos(\phi_1+\phi_2)$, and the event-by-event factor $\alpha^{z_1,z_2}_{M_1,M_2}$ is determined from the diFFs as
\begin{align}
    \alpha^{z_1,z_2}_{M_1,M_2}&= \frac{1}{2}  \frac{\sum_q e_q^2 H_1^{\sphericalangle,q}(z_1, M_1) H_1^{\sphericalangle,\bar{q}}(z_2, M_2)}{\sum_q e_q^2 D_1^q(z_1,M_1)D_1^{\bar{q}}(z_2,M_2)}.
\end{align}
Thus, the Bell variable is reconstructed from two parts, the experimental asymmetry observable $A_{12}$ that depends on the strength of transverse spin correlation of $q\bar q$, and the theory prediction of the spin analyzing power $\alpha^{z_1,z_2}_{M_1,M_2}$ that is independent of the $q\bar q$ spin state.

\begin{figure}
    \centering
    \includegraphics{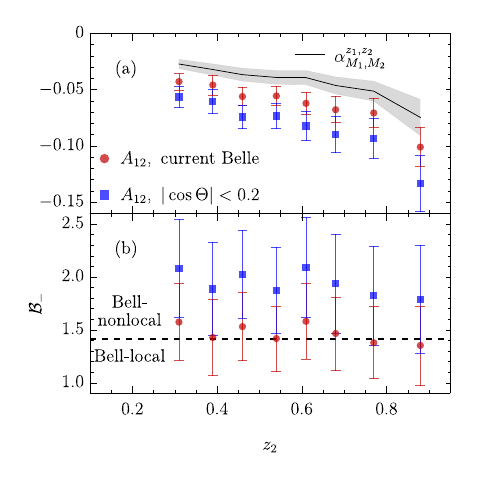}
    \caption{(a) The current measurement of $A_{12}$ (red points) and expected measurement of $A_{12}$ with a selection cut of $|\cos\Theta|<0.2$ (blue points), together with the spin analyzing power $\factorAlpha$ (black line) calculated with JAMDiFF~\cite{Cocuzza:2023oam,Cocuzza:2023vqs}. (b) The Bell variable $\mathcal{B}_-=A_{12}/\factorAlpha$, reconstructed from current data set (red) and expected data with a selection cut of $|\cos\Theta|<0.2$ (blue).}
    \label{fig:z1LastBin}
\end{figure}

\emph{Estimated Sensitivity.---}
With the method developed above, we revisit the existing data reported by Belle~\cite{Belle:2011cur}. This dataset consists of 81 data points representing the measured values of $A_{12}$, categorized into $9\times 9$ bins of $(z_1, z_2)$.
For the theory prediction of $\factorAlpha$, we utilize the JAMDiFF framework~\cite{Cocuzza:2023oam,Cocuzza:2023vqs} for the values of $D_1^q(z_i,M_i)$ and $H_1^{\sphericalangle,q}(z_i,M_i)$, along with their associated uncertainties. 
Given that the JAMDiFF framework exhibits limitations in effectively describing the small $z_i$ region~\cite{Cocuzza:2023vqs}, we exclude the initial bins of $z_1$ and $z_2$ from the Belle dataset~\cite{Belle:2011cur} where $z_i<0.275$. Subsequently, we focus on the remaining $64$ data points of $A_{12}$, corresponding to the $(z_1, z_2)$ bins ranging from $0.275$ to $1.0$. For each of these 64 data points, we calculate the spin analyzing power $\factorAlpha$ and reconstruct the Bell variable defined in Eq.~\eqref{eq:BellVariable}.  When calculating $\factorAlpha$, we sum over four flavors $(u,d,s,c)$ to be consistent with the current dataset.
The uncertainty associated with $\mathcal{B}_-$ comes from two sources: the experimental uncertainty of $A_{12}$ and the theoretical uncertainty of the diFFs,
\begin{equation}
    (\delta \mathcal{B}_-)^2 = 
    \left|\frac{\delta A_{12}}{\alpha^{z_1,z_2}_{M_1,M_2}} \right|^2 + 
    \left|\frac{A_{12}~ \delta \factorAlpha }{(\alpha^{z_1,z_2}_{M_1,M_2})^2 } \right|^2.
\end{equation}
As an illustrative example,  in Fig.~\ref{fig:z1LastBin} we consider the 8 data points in the last bin of $z_1$ and depict the measured value of $A_{12}$ and the reconstructed value of $\mathcal{B}_-$ with red dots.
It shows that the Bell variables $\mathcal{B}_-$ reconstructed from the current data hover near the boundary of Bell locality, exhibiting no significant evidence of Bell inequality violation. This observation can be attributed to the absence of optimal cuts applied to the current data to identify the most entangled region.

To further select the Bell state, we consider a selection cut of $|\cos\Theta|<0.2$ and show the expected value of $A_{12}$ and $\mathcal{B}_{-}$ with the blue points in Fig.~\ref{fig:z1LastBin}, where a clearer signal of Bell inequality violation can be found. The expected value of $A_{12}$ is estimated by scaling the measured value of $A_{12}$ under the leading order SM prediction of the $q\bar q$ spin correlation:
\begin{equation}
    \frac{A_{12}^{|\cos\Theta|<c_a}}{A_{12}^{|\cos\Theta|<c_b}} = \frac{3-c_a^2}{3+c_a^2} \frac{3+c_b^2}{3-c_b^2}.
\end{equation}
Both the statistical and systematic uncertainties from the experimental data~\cite{Belle:2011cur} are accounted for. We rescale the statistical uncertainty according to $1/\sqrt{N}$, with event number $N$ scaling as
\begin{equation}
    \frac{N^{|\cos\Theta|<c_a}}{N^{|\cos\Theta|<c_b}} = \frac{c_a(3+c_a^2)}{c_b(3+c_b^2)},
\end{equation}
while the systematic uncertainty is assumed to remain the same with the current experimental analysis after the rescaling. For the current Belle data, the expected value of the thrust angle aligns with a parton-level cut of $|\cos\Theta| < 0.67$ at leading order, and we scale the number of events and $A_{12}$ accordingly.

\begin{figure}
    \centering
    \includegraphics{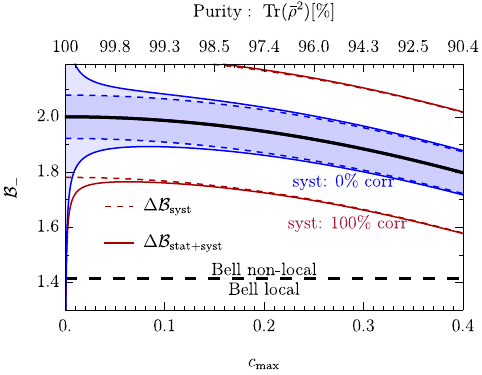}
    \caption{Expected result of $\mathcal{B}_-\pm \Delta \mathcal{B}$ as a function of selection cut $|\cos\Theta|<c_{\rm max}$, assuming the systematical uncertainties in each bin are uncorrelated (blue) or 100\% correlated (red).}
    \label{fig:Bccut}
\end{figure}

We further consider different selection cuts with $|\cos\Theta|<c_{\rm max}$ and present the combined results of $\mathcal{B}_-$ as a function of $c_{\rm max}$ in Fig.~\ref{fig:Bccut}.  Under factorization, the data points obtained in various phase spaces describe the identical $q\bar q$ spin state $\bar\rho$. Therefore, we present the combined results of the 64 Bell variables weighted by number of events,
\begin{equation}
    \mathcal{B}_- = \frac{1}{\sum_i N_i}\sum_i N_i \mathcal{B}_{-,i},
\end{equation}
where $i$ sums over all the 64 data points we consider, and $N_i \propto (1/\delta A_{12}^{\rm stat})^2$ is the number of events in the $i$'th bin.
The variance of the combined Bell variable is given by
\begin{equation}
    \mathrm{Var}( \mathcal{B}_-) = \sum_i  \frac{N_i^2}{N^2} \mathrm{Var} (\mathcal{B}_{-,i}) + \sum_{i\neq j} \frac{N_i N_j}{N^2} \mathrm{Cov}(\mathcal{B}_{-,i},\mathcal{B}_{-,j}). 
\end{equation}
Since the covariance matrix is not explicitly provided in Ref.~\cite{Belle:2011cur}, we assume the systematic uncertainties are either 0\% or 100\% correlated, while both the statistical uncertainties and the theory uncertainties are considered uncorrelated.  The expected results of $\mathcal{B}_-$ under both assumptions are depicted in Fig.~\ref{fig:Bccut}.
Given the significant number of events collected in Belle ($\sim 10^8$), the statistical uncertainty is much smaller than the systematic uncertainty.  Consequently, imposing a more stringent cut does not have much effect on the uncertainty, but yields a purer Bell state with a larger signal of Bell inequality violation.
By selecting an angular cut of $|\cos\Theta|<0.1$ above the jet resolution~\cite{Belle:2000cnh}, one can analyze a highly pure Bell state with $\text{Tr}(\bar{\rho}^2)>0.99$ and anticipate detecting Bell inequality violation with a sensitivity of 2.5 $\sigma$, even in the worst-case scenario where the systematic uncertainties in each bin are 100\% correlated.  In the case of assuming uncorrelated systematic uncertainties, we expect to achieve a sensitivity of 6.2 $\sigma$.

\emph{Conclusions and Discussions.---}
The collider environment is appealing for quantum information studies with many highly demanding quantum measurements naturally implemented, inspiring both fields of high-energy physics and quantum information~\cite{Fabbrichesi:2022ovb,Aoude:2022imd,Altakach:2022ywa,Fabbrichesi:2023jep,Bernal:2023ruk,Aoude:2023hxv,Maltoni:2024tul,Du:2024sly,Fabbrichesi:2025ywl}. Of the rich quantum systems produced at colliders, previous studies are mostly limited to massive particles with perturbative decay. Establishing the framework of studying quantum information for light quarks would make vast existing data available and allow us to make better use of the potential of colliders for quantum information study. 

We propose a novel approach to investigate the Bell inequality violation of the massless quark pairs by utilizing the interference diFF in $\pi^+\pi^-$ dihadron pair production at lepton colliders.  We demonstrated that the Bell inequality of quark pair system can be measured with a single observable, the azimuthal angle correlation of the dihadron pairs.
Our method represents the first examination of Bell inequality violation in massless quark pairs and the first exploration of a highly pure spin Bell state produced at colliders.
By implementing a straightforward additional selection cut on the hard scattering angle, we demonstrated that the measurement of Bell inequality violation can achieve a significance of 2.5 $\sigma$ in the worst-case scenario of 100\% correlated systematic uncertainties for the current Belle experiment, while the significance substantially exceeds $5\sigma$ when considering uncorrelated systematic uncertainties. Therefore, by comprehensively addressing systematic uncertainties, it becomes feasible to confirm the violation of the Bell inequality using available data, possibly resulting in the first measurement of Bell inequality violation in a collider scattering process. The detection of Bell inequality violation would also imply the presence of entanglement.

We examine dihadron pair production within the collinear factorization framework as a first example to investigate the Bell inequality violation of the massless quark pairs. With a more careful treatment, it is also possible to generalize our method within the TMD framework~\cite{Boer:2004mv,Metz:2016swz,Aidala:2021pvc}, which we leave as future works

\vspace{3mm}
\begin{acknowledgments}
We thank Chris Cocuzza for the help on the usage of JAMDiFF, and thank Matthew Low, Xin-Kai Wen and Zhite Yu for the helpful discussion.
Kun Cheng is supported in part by U.S. Department of Energy under grant No. DE-SC0007914.
Bin Yan is supported in part by the National Science Foundation of China under Grant No.~12422506, the IHEP under Grant No.~E25153U1 and CAS under Grant No.~E429A6M1. The authors gratefully acknowledge
the valuable discussions and insights provided by the
members of the Collaboration on Precision Tests and
New Physics (CPTNP).
\end{acknowledgments}

\bibliographystyle{apsrev4-1}
\bibliography{ref}

\end{document}